\documentclass[prl,aps,twocolumn,amsmath,amssymb,superscriptaddress]{revtex4}

\usepackage[english]{babel}
\usepackage{natbib}
\usepackage{amsfonts}
\usepackage{amssymb}
\usepackage{graphicx}
\usepackage{dcolumn}
\usepackage{amsmath}
\usepackage[ansinew]{inputenc}
\usepackage{psfig}
\usepackage{amssymb}

\begin{document}
\date{\today}
 \title
{Numerical simulation evidence of dynamical transverse Meissner effect and moving Bose glass phase}
\author{E. Olive}
\affiliation{LEMA, CNRS-UMR 6157, CEA, Universit\'e Fran\c{c}ois Rabelais, Parc de Grandmont 37200 Tours, France}
\author{J.C. Soret}
\affiliation{LEMA, CNRS-UMR 6157, CEA, Universit\'e Fran\c{c}ois Rabelais, Parc de Grandmont 37200 Tours, France}
\author{P. Le Doussal}
\affiliation{LPTENS, CNRS-UMR 8549, ENS, Universit\'e Paris-Sud, 24 rue Lhomond 75231 Cedex 05 Paris, France}
\author{T. Giamarchi}
\affiliation{DPMC, University of Geneva, 24 Quai Ernest Ansermet, 1211 Geneva 4, Switzerland}


\begin{abstract}
We present 3D numerical simulation results of moving vortex lattices in presence of 1D correlated disorder at zero temperature. 
Our results with field tilting confirm the theoritical predictions of a moving Bose glass phase, characterized by transverse pinning 
and dynamical transverse Meissner effect, the moving flux lines being localized along the correlated disorder direction.
Beyond a critical transverse field, vortex lines exhibit along all their length a "kink" structure resulting from 
an effective static "tin roof" pinning potential in the transverse direction. 
 \end{abstract}

\maketitle


 
Depinning and dynamical ordering of periodic lattices, such as vortex lattices (VL) driven by an external force in presence of static 
point disorder has been widely studied these last years. A large velocity expansion  \cite{Koshelev} showed that the motion reduces the effects of disorder.
In this approximation the effects of disorder on moving VL results only in an 
additional "shaking" temperature $T_{sh}\sim 1/v$ with expectation of moving crystalline VL at large velocities. 
It was however shown in \cite{Giamarchi} that, due to the components of the disorder and displacements transverse to the motion, a static disorder persists even
at large velocity, and leads to a moving glass phase (MG). The motion of particles in the MG occurs along rough static channels. Depending on the velocity the 
channels can be coupled or decoupled due to dislocations, thus giving two generic moving glasses. For weak disorder or large velocity in
$d=3$, channels are elastically coupled leading to a topologically ordered structure: the moving Bragg glass (MBG) \cite{Giamarchi} . For larger
disorder or in $d=2$ channels are decoupled by dislocations giving a moving transverse glass (MTG) or smectic \cite{Giamarchi, Balents} . The predicted channel
structure and dynamical ordering was confirmed by numerical \cite{Simul, Jensen, Gronbech, Fangohr} and  further analytical \cite{Scheidl} studies and observed
experimentally \cite{Marchevsky}.

Within the MG model \cite{Giamarchi}, the situation for correlated disorder such as heavy-ion columnar tracks has been recently investigated 
\cite{Chauve}. For all velocities at $T=0$ or for weak velocities at $T>0$ it was predicted a moving Bose glass phase (MBoG) characterized by a transverse critical force
and a diverging tilt modulus due to localization effect arising from the columns. This novel feature specific to correlated disorder results in a vanishing tilt response below a 
critical transverse field. This so-called dynamical transverse Meissner effect (DTME) is a crucial consequence of the MG theory; its observation should provide
an unambiguous signature of the MBoG phase. However, up to now there is no experimental nor numerical simulation evidence for this phase.
 
This Letter reports 3D numerical simulation of field tilting in a moving layered vortex system at zero temperature in a random pinning landscape of columnar defects. 
The elastically moving vortex lines confirms the existence of MBoG displaying DTME. Above a critical angle, the vortex lines exhibit a "kink" structure along 
their length. This results from the existence of an effective static pinning landscape which is almost periodic in the transverse direction to the vortex flow. 
Finite size effects are discussed within a simple model of a single flux line in a static periodic potential. In particular, it confirms the existence of a moving "kink" structure 
observed in our simulation.

We model a stack of $N_z$ Josephson-coupled parallel superconducting planes of thickness $d$ with interlayer spacing $s$. Each layer in the $(x,y)$ plane 
contains $N_v$ pancake vortices interacting with a random pinning background of 1D correlated disorder, namely $N_p$ 
columnar pins parallel to the $z$ direction. At $T=0$ the overdamped equation of motion of a pancake $i$ in position $\bold r_i(z)$ reads
$$\eta {{d{\bf r_i}}\over{dt}}=-{\sum_{j \neq i}}\nabla_i U^{vv}(\rho_{ij},z_{ij})-{\sum_{p}}\nabla_i U^{vp}(\rho_{ip})+{\bf F}^L+{\bf F}^{tilt}(z)$$
where $\rho_{ij}$ and $z_{ij}$ are the components of ${\bf r_{ij}}={\bf r_i}-{\bf r_j}$ in cylindrical coordinates, $ \rho_{ip}$
 is the in-plane distance between the pancake $i$ and a pinning site in the same layer at ${\bf r_p}$, and $\nabla_i$ is the 2D gradient operator acting in the 
$(x,y)$ plane. The viscosity coefficient is $\eta$, ${\bf F}^L=F^L{\bf \hat x}$ is the Lorentz driving force due to an applied current and 
${\bf F}^{tilt}(z)$ is the surface force due to the field tilting away from the $z$ axis  in the $y$ direction. This force acts as a torque on each flux line, {\it i.e.} 
${\bf F}^{tilt}(z=0)=-{\bf F}^{tilt}(z=N_zs)=F^{tilt}{\bf \hat y}$ and ${\bf F}^{tilt}(z)={\bf 0}$ for pancakes in the bulk. The tilting force modulus is defined by
$F^{tilt}=\epsilon^2\phi_0H_y/4\pi=8\pi\epsilon^2\epsilon_0\lambda_{ab}^2H_y/\sqrt 3 a_0^2H_z$, where $\epsilon_0=(\phi_0/4\pi\lambda_{ab})^2$, 
$\lambda_{ab}$ is the in-plane magnetic penetration depth, $a_0$ is the average vortex distance, $H_y$ is the transverse field component, and $\epsilon$ is the 
anisotropy parameter. 
The intra-plane vortex-vortex repulsive interaction is given by a modified Bessel  function $U^{vv}(\rho_{ij})=2\epsilon_0dK_0(\rho_{ij}/\lambda_{ab} )$. 
Following \cite{Ryu}, the inter-plane attractive interaction between pancakes in adjacent layers of altitude $z$ and $(z+s)$ reads 
$U^{vv}(\rho_{ij},z_{ij}=s)=(2s\epsilon_0/\pi)[1+ln(\lambda_{ab}/s)][(\rho_{ij}/2r_g)^2-1]$ for $\rho_{ij}\le 2r_g$ 
and $U^{vv}(\rho_{ij},z_{ij}=s)=(2s\epsilon_0/\pi)[1+ln(\lambda_{ab}/s)][\rho_{ij}/r_g-2]$ otherwise;  in this expression $r_g=\xi _{ab}/\epsilon $, 
where $\xi_{ab}$ is the in-plane coherence length. 
Finally, the attractive pinning potential is given by $U^{vp}( \rho_{ip})=-\alpha A_pe^{-(\rho_{ip}/R_p)^2}$, where  
$A_p=(\epsilon _0d/2)ln[1+(R_p^2/2\xi_{ab}^2)]$ as proposed in \cite{Mkrtchyan} and $\alpha$ is a tunable parameter. 
We consider periodic boundary conditions of $(L_x, L_y)$ sizes in the $(x,y)$ plane while open boundaries are taken in the $z$ direction. All details about our 
method for computing the Bessel potential with periodic conditions can be found in \cite{Olive}. Molecular Dynamics simulation is used for $N_v=30$ 
vortex lines in a rectangular basic cell $(L_x,L_y)=(5, 6\sqrt 3/2)\lambda_{ab}$, and for a number of layers varying from $N_z=19$ up to $N_z=149$. 
The number of columnar pins is set to $N_p=30$. We consider the London limit $\kappa =\lambda_{ab} /\xi_{ab} =90$, with an average vortex distance 
$a_0=\lambda_{ab} $, and  $d=2.83\ 10^{-3}\lambda_{ab}$,  $s=8.33\ 10^{-3}\lambda_{ab}$, $R_p=0.22\ \lambda_{ab}$, $\epsilon =0.01$, $\eta=1$. 
These values are coherent with strong anisotropy layered superconductors with $B_z \sim 1000\ G$.
The choice of the second order Runge-Kutta algorithm time iteration step $\delta t$ is dictated by the dominant force $F^L$, and we take $\delta t=7\ 10^{-2}t_0$, 
where $t_0=\eta\xi_{ab}/F^L$. We choose the tunable pinning parameter $\alpha =1/25$ so that the maximum pinning force is $F_{max}^{vp}\sim F_0/5$ 
where $F_0=2\epsilon _0d/\lambda_{ab}$ is the unit force defined by the Bessel interaction.
Finally, the driving force applied along a principal vortex lattice direction $x$ is set to $F^L\sim 2.8F_0$. This corresponds to the fully elastic flow limit since 
$F^L\sim 25F_c^L$ where $F_c^L$ is the critical Lorentz force along $x$. 

The "experimental" procedure is the folllowing: we start in 2D with the $(x,y)$ plane by randomly throwing $N_v$ pancakes and $N_p$ gaussian pins, 
and relaxation with zero Lorentz force yields a vortex structure with dislocations. The Lorentz force is then slowly increased far in the elastic phase up to 
$F^L\sim2.8F_0$ with steps $\delta F^L\sim3.5\ 10^{-3}F_0$ every $3.6\ 10^4t_0$. The successive dynamical regimes observed in our 2D vortex system 
are the following: (i) pinned regime where all vortices have zero velocity; (ii) plastic channels flowing through pinned regions and where the motion is periodic 
(washboard frequency) as seen in \cite{Jensen, Gronbech}; (iii) plastic turbulent flow with no stationnary vortices as seen in \cite{Gronbech, Fangohr}; 
(iv) weakly decoupled channels named MTG since quasi long range longitudinal order exists in each channel; (v) elastic phase where all vortices have the same 
average velocity and named MBG since no dislocation appears and the motion occurs through rough static channels even in the high driving phase 
$F^L\sim 25F_c^L$ studied below. Our 2D numerical results are therefore in agreement with the moving glass theory developed in \cite{Giamarchi}. 
Finally, note that they are consistent with the phase diagram region corresponding to the intermediate pinning of Ref \onlinecite{Fangohr}. 

\begin{figure}
\includegraphics[width=0.9\linewidth]{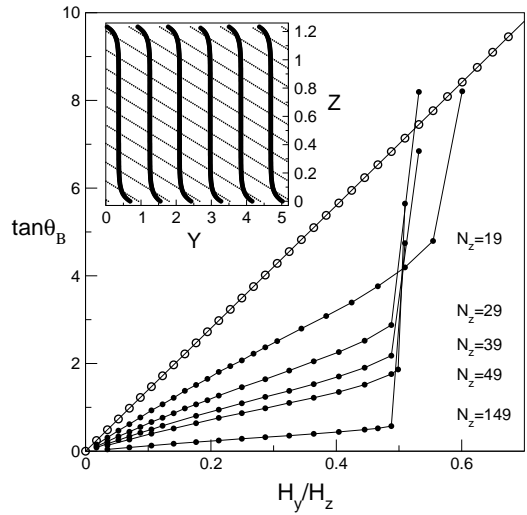}
\caption{ \label{fig1} Average vortex line inclination  $\tan \theta_B=B_y/B_z$ versus the field inclination $\tan \theta_H=H_y/H_z$ for several system sizes
$N_z=19, 29, 39, 49, 149$ (filled circles). The linear response of the pinning free vortex system is shown with open circles. Size effects show that the partial 
transverse screening observed for $H_y/H_z\prec 0.50$ tends to be total in the infinite size limit $N_z \rightarrow \infty $, showing the existence of the DTME 
below the critical transverse field $H_y^c/H_z\sim 0.50$. 
Insert: projection in the $(y,z)$ plane of the 30 vortex lines composed of $N_z=149$ layers illustrating the MBoG phase (thick lines) and DTME, compared to 
the vortex lines of the non-disordered system (thin oblique lines). In the MBoG phase, the transverse field penetrates only at the surface and vortex lines in the bulk are 
aligned with the columnar defect direction $z$. In the pinning free situation, the vortex lines form straight lines in an oblique direction defined by the field inclination 
and Josephson coupling. Note that the whole vortex lattice is moving elastically in the $x$ direction, {\it i.e.} perpendicularly to the image.}
\end{figure}

Now starting from this situation, we extend the vortices and the pins in the $z$ direction in order to obtain 3D straight vortex lines moving elastically along the 
$x$ direction through a random columnar pin landscape. Then, starting from this situation, we now slowly tilt the magnetic field ${\bf H}$ away from $z$ in the 
$y$ direction by increasing the $H_y$ component transversally to the vortex flow. For several system sizes, Fig.\ \ref{fig1} displays the vortex line response to low 
field tilting, {\it i.e.} the average vortex line inclination $\tan \theta_B=B_y/B_z$ versus the field inclination $\tan \theta_H=H_y/H_z$, compared with 
the linear response of the pinning free vortex system. It shows a partial screening of the transverse component $H_y$ of the field, which tends to be 
total for infinite vortex line length since the slopes tend towards zero when $N_z \rightarrow \infty $. This transverse screening is followed by a transition at a critical 
field $H_y^c/H_z\sim 0.50$, above which the average vortex line inclination recovers the one obtained without any pinning. 
Therefore, Fig.\ \ref{fig1} shows the existence of DTME which is the very signature of the MBoG phase as predicted in \cite{Chauve}. As a good illustration of 
DTME, we show in the insert of Fig.\ \ref{fig1}, the vortex line configuration of the MBoG compared with the non-disordered lattice obtained for the same field tilting.
One clearly sees that below the transition the transverse field 
enters at the surface without penetrating into the bulk, in opposition to the pinning free vortex system. So below the critical transverse field, the vortex lines remain 
parallel to the columnar defects even though they are moving at enough large velocity.

\begin{figure}
\includegraphics[width=0.9\linewidth]{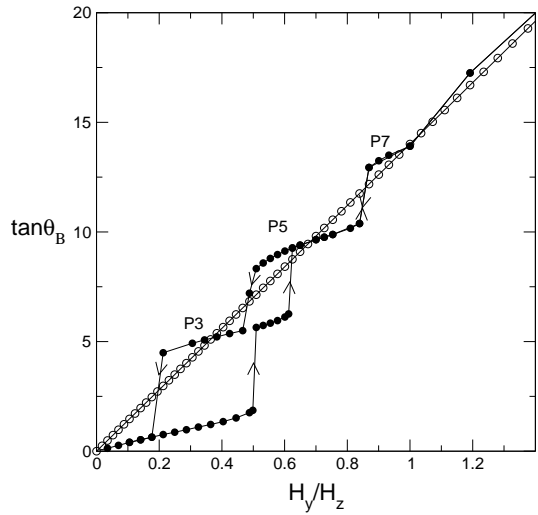}
\caption{ \label{fig2}Plateaux and hysteresis observed above the transition for $N_z=49$. The origin of each plateau is connected to a given number of "pinned" 
regions passed through by each vortex line. For details about the plateaux labels and the "pinned" regions defining the effective static pinning landscape, see Fig.\ \ref{fig3} and text below.
Finite size effects show that the plateaux and hysteresis disappear in the infinite size limit $N_z \rightarrow \infty $. Plateaux and hysteresis may therefore be observed 
in thin real samples.
} 
\end{figure}

\begin{figure}
\includegraphics[width=0.9\linewidth]{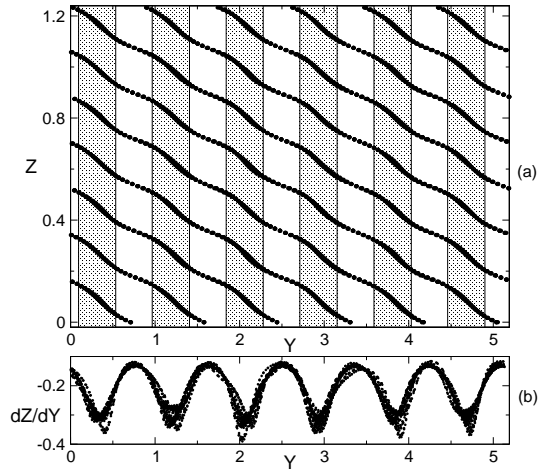}
\caption{ \label{fig3}(a) Projection in the $(y,z)$ plane of the 30 moving vortex line composed of $N_z=149$ layers and obtained for $H_y/H_z\sim 0.53$ while 
increasing the transverse field. In grey are materialised the "pinned" regions which define an effective pinning landscape which is static and almost periodic.
(b) Derivative $dz/dy$ for all the 30 vortex lines shown in (a). The vortex line modulations along their length become clearly visible. Taking into account the minima 
and inflexion points of the derivative allows the determination of the "pinned" regions.} 
\end{figure}

We shall now examine the situation above the critical transverse field. Far above $H_y^c/H_z$ and whatever the vortex line size is, the average vortex line inclination
recovers the one obtained without any pinning (not shown). A more interesting case happens in the intermediate transverse field range above the transition
and shown for $N_z=49$ layers in Fig.\ \ref{fig2}. Increasing the transverse field above its critical value $H_y^c/H_z\sim 0.50$ gives rise to plateaux in the 
$\tan\theta_B$ vs $H_y/H_z$ curve.
A consequence of these plateaux is that decreasing the transverse field generates hysteresis in the vortex inclination. A typical vortex line configuration above the 
transition is shown for $N_z=149$ layers in Fig.\ \ref{fig3}(a). A modulated structure $z(y)$ appears along their length.
This modulation is more obvious looking at the derivative $dz/dy$ shown for all the vortex lines in Fig.\ \ref{fig3}(b). By using the minima and maxima of the derivative 
we may roughly define two parts of small and big inclinations in the vortex lines with respect to $z$. The small ones are termed "pinned" regions and are linked by 
"unpinned" regions where the vortex inclination is bigger. The materialisation of the "pinned"  regions is shown in grey in Fig.\ \ref{fig3}(a). We visualize in this way 
a static pinning
landscape in the transverse direction $y$ with effective columnar pin width corresponding to twice the input pinning "radius" $R_p$. Each plateau of Fig.\ \ref{fig2} 
can be thereby labelled with an odd number corresponding to the
number of "pinned" regions passed through by each vortex line. In this way, while increasing the transverse field above the transition, 
the vortex lines firstly pass through 3 pinned regions for a given range of $H_y$ defining the "P3" plateau, then the system jumps to the following
plateau "P5" where the vortex lines pass through 5 pinned regions, and so on.
The same process holds while decreasing the transverse field generating thereby hysteresis. 
Finally, the central point displayed in Fig.\ \ref{fig3}(a) is that the vortex lines which are moving in the $x$ direction perpendicularly to the 
figure plane see an effective static tin roof potential in the transverse direction $y$. This is a direct consequence of the existence of static pinned channels of motion
\cite{Giamarchi, Chauve}, and clearly shows that the action of the disorder in the moving system cannot be simply reduced \cite{Koshelev} to a temperature.
Note that this effective potential is not strictly periodic because of the channel roughness. 

\begin{figure}
\includegraphics[width=0.9\linewidth]{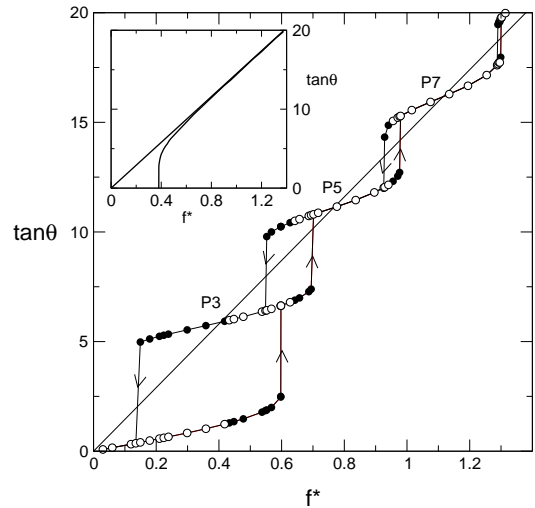}
\caption{ \label{fig4}Plateaux and hysteresis observed within the simple model (Eq. \ref{eq1}) for a finite length $L=0.4\ a_0$ and $g/c=18.75$. Black filled circles 
represent the metastable states followed by the elastic line when the surface force $f$  is slowly increased, then decreased. The white filled circles represent the true 
equilibrium states.
Insert: Eq. \ref{eq1} results in the infinite size limit $L\rightarrow\infty $ showing the existence of a critical dimensionless force 
$f^*_c=\sqrt 3/2\pi^2.\sqrt{g/c}$
and the disappearance of the plateaux and hysteresis.} 
\end{figure}

We now turn to the discussion of the finite size effects in our simulation by considering the tilt response of a single static elastic vortex line in a tin roof pinning
potential at $T=0$. Such a simple model allows to get a valuable insight into the transverse properties of the moving vortex lattice in presence of correlated disorder,
{\it e.g.} the TME and the kink vortex line structure. The energy of a line of length $L$ is given by:
\begin{eqnarray}  
\label{eq1}
E(u)=\int_{0}^{L}dz\left ({c \over 2}\left ({du \over dz}\right )^2+V_{eff}(u)\right)+f\left(u(L)-u(0)\right),
\end{eqnarray} 
where $u(z)$ is the one-dimensional displacement field in the $y$ direction, $c=\epsilon^2\epsilon_0$ is the elastic constant, $f\propto H_y$ is a surface force, and
$V_{eff}(u)=g\left[1-\cos(4\pi u/\sqrt3a_0)\right]$ with $g$ being a constant. There exists a ground state $u_{L,f}$ which minimizes $E(u)$. We calculate the tilt response
of the vortex line defined by $\tan\theta=(u_{L,f}(L)-u_{L,f}(0))/L$. In the limit $L\rightarrow\infty $, we find a critical force $f_c=4\sqrt{gc}/\pi$ below which
$u_{L=\infty,f}\equiv 0$, giving $\tan\theta=0$ and hence TME. For $f \succ f_c$ the vortex line is tilted and its ground state $u_{L=\infty,f}(z)$ are expressed in terms of
Jacobian elliptic functions which generate kink vortex line structures as observed in our simulation (see Fig.\ \ref{fig3}(a)). The tilt response is shown in the insert of Fig.\ \ref{fig4} together with the pure
elastic response, {\it i.e.} $g=0$. In the latter case, the ground states are linear functions and hence the vortex is a straight line of slope $\tan\theta=f/c$. The kink vortex
line structure obviously rely only on the periodic potential. Note that $f^*=\sqrt 3 f/8\pi c$ is the equivalent quantity of $H_y/H_z$. For a finite size line and
$f=0$, the minimum energy solution is $u_{L,f=0}\equiv 0$, {\it i.e.} the line is pinned. If $f$ is nonzero but sufficiently weak, both extremities of the line are pulled away 
from its minimum energy configuration in a continuous way, yielding a nonzero tilt response as observed in our simulation (see Fig.\ \ref{fig1} and insert). 
Furthermore, for given values of $f$, the vortex line abruptly stretches to
a more favourable position resulting in jumps in the tilt response. We show in Fig.\ \ref{fig4} the tilt response computed for $L=0.4\ a_0$ which is the equivalent size
of  the 49 layers in our simulation results shown in Fig.\ \ref{fig2}; the straight line displays the pure elatic response $g=0$. We therefore find plateaux and hysteresis
which are very close to the ones observed in our simulation and shown in Fig.\ \ref{fig2}.

In conclusion, our numerical simulation results give the first confirmation of the existence of the MBoG phase displaying a DTME developed in \cite{Chauve}.
Furthermore, above the critical transverse field, the simulation revealed a "kinked" vortex line structure resulting from an almost periodic effective static pinning 
potential in the transverse direction.

\begin{acknowledgments}
We are grateful to the Laboratoire de Mathématiques et Physique Théorique -
CNRS UMR 6083 Université F. Rabelais - Tours (France) for our extensive use of their computers.
\end{acknowledgments}



\references
\bibitem{Koshelev} A. E. Koshelev, V. M. Vinokur, \prl {\bf 73}, 3580 (1994).
\bibitem{Giamarchi} T. Giamarchi, P. Le Doussal, \prl {\bf 76}, 3408 (1996); 
{\it ibid.} {\bf 78}, 752 (1997);
P. Le Doussal, T. Giamarchi, \prb {\bf 57}, 11356 (1998).
\bibitem{Balents} L. Balents, M. C. Marchetti, L. Radzihovsky, \prl {\bf 78}, 751 (1997);  
\prb {\bf 57}, 7705 (1998).
\bibitem{Simul} K. Moon, R. T. Scalettar, G. T. Zimanyi, \prl {\bf 77}, 2778 (1996); 
S. Ryu {\it et al.}, \prl {\bf 77}, 5114 (1996); 
S. Spencer, H. J. Jensen,  \prb {\bf 55}, 8473 (1997); 
C. J. Olson, C. Reichhardt, F. Nori,  \prl {\bf 81}, 3757 (1998); 
A. B. Kolton, D. Dominguez, C. J. Olson, N. Gronbech-Jensen, \prb {\bf 62}, R14657 (2000).
\bibitem{Jensen} H. J. Jensen, A. Brass, A. J. Berlinsky, \prl {\bf 60}, 1676 (1988); 
\bibitem{Gronbech} 	N. Gronbech-Jensen, A. R. Bishop, D. Dominguez, \prl {\bf 76}, 2985 (1996).
\bibitem{Fangohr}  H. Fangohr, S. J. Cox, P. A. J. de Groot, \prb {\bf 64}, 064505 (2001).
\bibitem{Scheidl} S. Scheidl, V. M. Vinokur, \pre {\bf 57}, 2574 (1998); 
\bibitem{Marchevsky} M. Marchevsky, J. Aarts, P. H. Kes, M. V. Indenbom, \prl {\bf 78}, 531 (1997); 
A. M. Troyanovski, J. Aarts, P. H. Kes, Nature (London) {\bf 399}, 665 (1999); 
S. Bhattacharya, M. J. Higgins, \prl {\bf 70}, 2617 (1993);
U. Yaron {\it et al.}, {\it ibid.} {\bf 73}, 2748 (1994); 
M. C. Hellerqvist {\it et al.}, {\it ibid.} {\bf 76}, 4022 (1996);
\bibitem{Chauve} 	P. Chauve, P. Le Doussal, T. Giamarchi, \prb {\bf 61}, R11906 (2000).
\bibitem{Ryu} 	S. Ryu, S. Doniach, G. Deutscher, A. Kapitulnik, \prl {\bf 68}, 710 (1992).
\bibitem{Mkrtchyan} G. S. Mkrtchyan, V. V. Schmidt, Sov. Phys. JETP {\bf 34}, 195; 
G. Blatter et al., \rmp {\bf 66}, 1125 (1994).
\bibitem{Olive} 	E. Olive, E.H. Brandt, \prb {\bf 57}, 13861 (1998).
\end{document}